\documentclass[fleqn,usenatbib]{mnras}
\usepackage[utf8]{inputenc}
\usepackage{graphicx}
\usepackage{hyperref}
\usepackage{amssymb}
\usepackage{amsmath}
\usepackage{amsfonts}

\title[(Sub-)Halo Mass Function on Sub-Galactic Scales]{The Normalization and Slope of the Dark Matter (Sub-)Halo Mass Function on Sub-Galactic Scales}
\author[Andrew J. Benson]{Andrew J. Benson\,$^{1}$\thanks{E-mail: abenson@carnegiescience.edu}\\
$^{1}$ Carnegie Observatories, 813 Santa Barbara Street, Pasadena, CA 91101, USA\\
}

\begin{document}

\maketitle

\begin{abstract}
 Simulations of cold dark matter make robust predictions about the slope and normalization of the dark matter halo and subhalo mass functions on small scales. Recent observational advances utilizing strong gravitational lensing have demonstrated the ability of this technique to place constraints on these quantities on subgalactic scales corresponding to dark matter halo masses of $10^6$--$10^9\mathrm{M}_\odot$. On these scales the physics of baryons, which make up around 17\% of the matter content of the Universe but which are not included in pure dark matter N-body simulations, are expected to affect the growth of structure and the collapse of dark matter halos. In this work we develop a semi-analytic model to predict the amplitude and slope of the dark matter halo and subhalo mass functions on subgalactic scales in the presence of baryons. We find that the halo mass function is suppressed by up to 25\%, and the slope is modified, ranging from $-1.916$ to $-1.868$ in this mass range. These results are consistent with current measurements, but differ sufficiently from the expectations for a dark matter only universe that it may be testable in the near future. 
\end{abstract}


\section{Introduction}

Recent work by \protect\cite{vegetti_inference_2014}, \protect\cite{hsueh_sharp_2019}, and \cite{gilman_warm_2019} has demonstrated that strong gravitational lensing can provide constraints on the normalization of the dark matter subhalo mass function on mass scales of $10^6$--$10^9\mathrm{M}_\odot$, with the possibility that future measurements may also constrain the slope of that mass function. There are robust predictions from cold dark matter (CDM) theory based largely on N-body simulations of pure dark matter universes for the slope of the mass function, and subhalo mass function, in this regime. For example, \cite{springel_aquarius_2008} find a best fit slope for the subhalo mass function of $\mathrm{d}\log N/\mathrm{d}\log m=-1.9$ in the Aquarius N-body simulations (varying between $-1.87$ and $-1.93$ depending on the exact mass range used in their fit), and \cite{fiacconi_cold_2016} finding a slope of $-1.877$ in the Ponos simulations. However, the mass scales of $10^6$--$10^9\mathrm{M}_\odot$ probed by strong gravitational lensing are comparable to the Jeans mass in the IGM post-reionization. For example, \cite{gnedin_effect_2000} show that the Jeans mass at mean density can reach over $10^{10}\mathrm{M}_\odot$ in the post-reionization universe.

Consequently, baryons on these scales do not act as a collisionless fluid and the growth of perturbations on these scales can no longer be treated using collisionless dynamics (e.g. with pure N-body simulation techniques). Instead, the hydrodynamics of the baryonic component must be taken into account.

These baryonic effects have been investigated using hydrodynamical simulations of structure formation. For example, \citeauthor{schaller_effects_2015}~(\citeyear{schaller_effects_2015}; their \S4.3.2, fig.~4.2) show the $z=0$ mass function of halos in the Eagle simulations compared to an equivalent dark matter only model. They find that the halo mass function is suppressed in the presence of baryons by around 30\% at $10^8\mathrm{M}_\odot$, with the suppression becoming consistent with 0\% above around $10^{12}\mathrm{M}_\odot$. \cite{qin_dark-ages_2017} examine the effects of baryons on halo formation using the DRAGONS simulations, exploring several models including an adiabatic model with no atomic cooling, stellar physics or reionization, through to models including gas cooling, star formation, and feedback. In their adiabatic model they find the mass function is suppressed by up to 60\% at $z=9$ at halo masses of $10^{10}\mathrm{M}_\odot$, while for halo masses of $10^8\mathrm{M}_\odot$ the suppression varies between 10--50\% between $z=2$ and $z=13$. The inclusion of reionization and feedback in their models leads to further suppression.

These prior works have not explicitly assessed the effects on the slope of the halo and subhalo mass functions, nor have they been able to probe the full range of halo masses relevant to strong gravitational lensing observations. Here, we develop a simple treatment for the nonlinear collapse of halos on scales below the Jeans mass. Our goal is to develop intuition for the magnitude of these effects, provide a framework for rapidly modeling these effects as a function of redshift and dark matter microphysics, and use this framework to estimate how the slope and normalization of the halo and subhalo mass functions are changed from the pure dark matter expectation on mass scales relevant to strong gravitational lensing constraints.

Throughout this work we adopt a cosmological model characterized by ($\Omega_\mathrm{m},\Omega_\mathrm{b},\Omega_\Lambda,H_0/\hbox{km}\,\hbox{s}^{-1} \hbox{Mpc}^{-1},\sigma_8,n_\mathrm{s}=0.275, 0.0458, 0.725, 70.2, 0.816, 0.968$; \citealt{komatsu_seven-year_2011}) for consistency with \cite{qin_dark-ages_2017} to which we make some comparisons. The framework presented in this work is not explicitly calibrated to this particular set of cosmological parameters however, and so can also be applied to universes characterized by other, more recent determinations of cosmological parameters.

\section{Methods}

To model the formation of halos we utilize the well-established framework of the spherical collapse model (\citealt{gunn_infall_1972,peebles_large-scale_1980}, \S19) for halo formation, together with Press-Schechter/excursion set techniques to model halo and subhalo mass functions \citep{press_formation_1974,bond_excursion_1991,bower_evolution_1991}. In the following subsections we generalize these models to scales below the Jeans mass in the IGM where the baryonic component no longer undergoes gravitational collapse.

\subsection{Linear Growth}

The coupled equations for the growth of linear perturbations in the dark matter and baryons are given by \citep{gnedin_probing_1998}:
\begin{eqnarray}
 \ddot{\delta}_\mathrm{X} + 2 H \dot{\delta}_\mathrm{X} &=& 4 \pi \mathrm{G} \bar{\rho} ( f_\mathrm{X} \delta_\mathrm{X} + f_\mathrm{b} \delta_\mathrm{b}) \nonumber \\
 \ddot{\delta}_\mathrm{b} + 2 H \dot{\delta}_\mathrm{b} &=& 4 \pi \mathrm{G} \bar{\rho} ( f_\mathrm{X} \delta_\mathrm{X} + f_\mathrm{b} \delta_\mathrm{b}) - {c_\mathrm{s}^2 \over a^2} k^2 \delta_\mathrm{b},
 \label{eq:linearGrowth}
\end{eqnarray}
where $\delta_\mathrm{X}$ and $\delta_\mathrm{b}$ are the overdensities of the dark matter and baryonic perturbations respectively, $H$ is the Hubble parameter, $\bar{\rho}$ is the mean density of the universe, $f_\mathrm{X}$ and $f_\mathrm{b}$ are the fractions of that mean density in the form of dark matter and baryons respectively, $c_\mathrm{s}$ is the sound speed in the baryons, $k$ is the comoving wavenumber of the perturbation, $a$ is the expansion factor, and an over dot indicates a derivative with respect to time.

In the limit of $k\rightarrow\infty$ perturbations in the baryonic component do not grow, and so $\delta_\mathrm{b}=0$ (after any possible initial perturbation has decayed away). In an Einstein-de Sitter cosmology (and, therefore, during the matter-dominated phase of general cosmologies) the resulting equation for the dark matter then admits power-law solutions in time, $t$, of the form $\delta_\mathrm{X} \propto t^p \propto a^{3p/2}$, where \citep{hu_small-scale_1998}:
\begin{equation}
 p(f_\mathrm{X}) = {\pm \sqrt{1 + 24 f_\mathrm{X}} - 1 \over 6}.
 \label{eq:growthExponent}
\end{equation}
Considering only the growing modes, in the limit of $f_\mathrm{X} \rightarrow 1$ we find $p=2/3$, or $\delta_\mathrm{X}\propto a$---the usual growth factor for a pure dark matter Einstein-de Sitter universe. For $f_\mathrm{X}<1$, $p<2/3$, so perturbations in the dark matter grow more slowly due to the presence of the non-clustering baryons. In the limit $f_\mathrm{X}\rightarrow 0$ we find $p\rightarrow 0$, and perturbations in the dark matter no longer grow (and will in fact decay away as the decaying mode has $p=-1/3$ for $f_\mathrm{X}=0$).

For the case of a general cosmological model and arbitrary wavenumber we use equations~(\ref{eq:linearGrowth}) to evolve linear perturbations in the post-recombination universe. We use CAMB \citep[][version 1.0.7]{challinor_linear_2011} to compute the transfer function to $z_\mathrm{i}=150$ to set the initial conditions $(\delta_\mathrm{X},\dot{\delta}_\mathrm{X},\delta_\mathrm{b},\dot{\delta}_\mathrm{b})$, and then integrate forward in time to solve for the growth of linear perturbations during the remainder of cosmic history. This initial redshift is chosen to be low enough such that the transfer function would be (almost) independent of time if the universe evolved adiabatically with no further heating, but high enough that no sources of heating (stars or AGN) have yet begun to form.

\subsection{Nonlinear Collapse}

We next consider the nonlinear collapse of spherical top-hat perturbations on scales $k\rightarrow \infty$ such that the baryonic component does not undergo gravitational collapse.

\subsubsection{Critical Overdensities}

A key ingredient in the family of Press-Schechter theories is the extrapolated linear theory overdensity at which a spherical perturbation collapses to zero radius, $\delta_\mathrm{c}$. Since for $k\rightarrow \infty$ the baryons do not undergo gravitational collapse, the energy within a spherical perturbation is not conserved \citep{weinberg_constraining_2003}, and the usual approach \citep[e.g.][]{percival_analytic_2000} used to solve for the critical overdensity for collapse can not be applied. Instead we must use the approach developed for the case of dark energy (for which the energy of the perturbation is similarly not conserved---e.g. \citealt{percival_cosmological_2005}, to which the reader is referred for a complete description of the approach). Briefly, the evolution of the perturbation radius, $a_\mathrm{p}$, is described by the cosmology equation for the perturbation, along with the cosmology and Friedmann equations for the background (which have identical solutions as the pure dark matter case since on the largest scales the pressure of the baryons has no effect):
\begin{eqnarray}
 {1 \over a_\mathrm{p}} {\mathrm{d}^2 a_\mathrm{p} \over \mathrm{d}(H t)^2} &=& -{1\over 2}\left[ \Omega_\mathrm{X} a_\mathrm{p}^{-3} + \Omega_\mathrm{b} a^{-3} - 2 \Omega_\Lambda \right], \nonumber \\
 {1 \over a} {\mathrm{d}^2 a \over \mathrm{d}(H t)^2} &=& -{1\over 2}\left[ \Omega_\mathrm{X} a^{-3} + \Omega_\mathrm{b} a^{-3} - 2 \Omega_\Lambda \right], \nonumber \\
 {1 \over a^2} \left[ {\mathrm{d} a \over \mathrm{d}(H t)}\right]^2 &=& \Omega_\mathrm{X} a^{-3} + \Omega_\mathrm{b} a^{-3} + \Omega_\mathrm{K} a^{-2} + \Omega_\Lambda ,
\end{eqnarray}
where $a_\mathrm{p}(t)$ is the radius of the spherical perturbation, $a(t)$ is the expansion factor of the Universe, $\Omega_\mathrm{X}$, $\Omega_\mathrm{b}$, and $\Omega_\Lambda$ are the density parameters for dark matter, baryons, and dark energy\footnote{Note that even though we use a subscript ``$\Lambda$'' our derivations follow \protect\cite{percival_cosmological_2005} in assuming a dark energy model with equation of state $w(a) = w_0 + w_1 a (1-a)$. The equations must be solved numerically even in the case of a cosmological constant, so there is no loss in simplicity by allowing for general dark energy models.} respectively.

\cite{percival_cosmological_2005} set the initial conditions for these equations using an analytic solution for $a_\mathrm{p}(t)$ appropriate in the limit $H t \rightarrow 0$ (in which case the $\Omega_\Lambda$ term can be neglected). However, in the case considered here no such analytic solution is available as the $\Omega_\mathrm{b}$ term is not negligible at early times. Instead, we proceed by directly choosing a small initial perturbation, $\delta_0$, at some early time, $t_0$, such that this perturbation is well within the linear regime. We further set the initial growth rate of this perturbation such that it matches the linear growth factor expected during the matter-dominated regime (see eqn.~\ref{eq:growthExponent}), i.e.
\begin{equation}
 {\dot{\delta}_0 \over \delta_0} = {\dot{D}(t_0) \over D(t_0)}.
\end{equation}
The initial radius of the perturbation is then found from the relation
\begin{equation}
 1+\delta_0 = \left[ {a(t_0) \over a_\mathrm{p}(t_0)} \right]^3,
\end{equation}
while the initial growth rate of the radius is found from the derivative of this equation
\begin{equation}
{1\over 3} {\dot{\delta}_0 \over 1 + \delta_0} =  {\dot{a} \over a} - {\dot{a}_\mathrm{p} \over a_\mathrm{p}}.
\end{equation}
For a given time, $t$, we seek the value of $\delta_0$ which results in collapse (i.e. $a_\mathrm{p}(t)=0$) at that time. The critical overdensity for collapse, extrapolated to the present day (as is conventional) is then simply $\delta_\mathrm{c}(t) = \delta_0 / D(t_0)$.

\begin{figure}
 \begin{tabular}{c}
 \includegraphics[width=80mm]{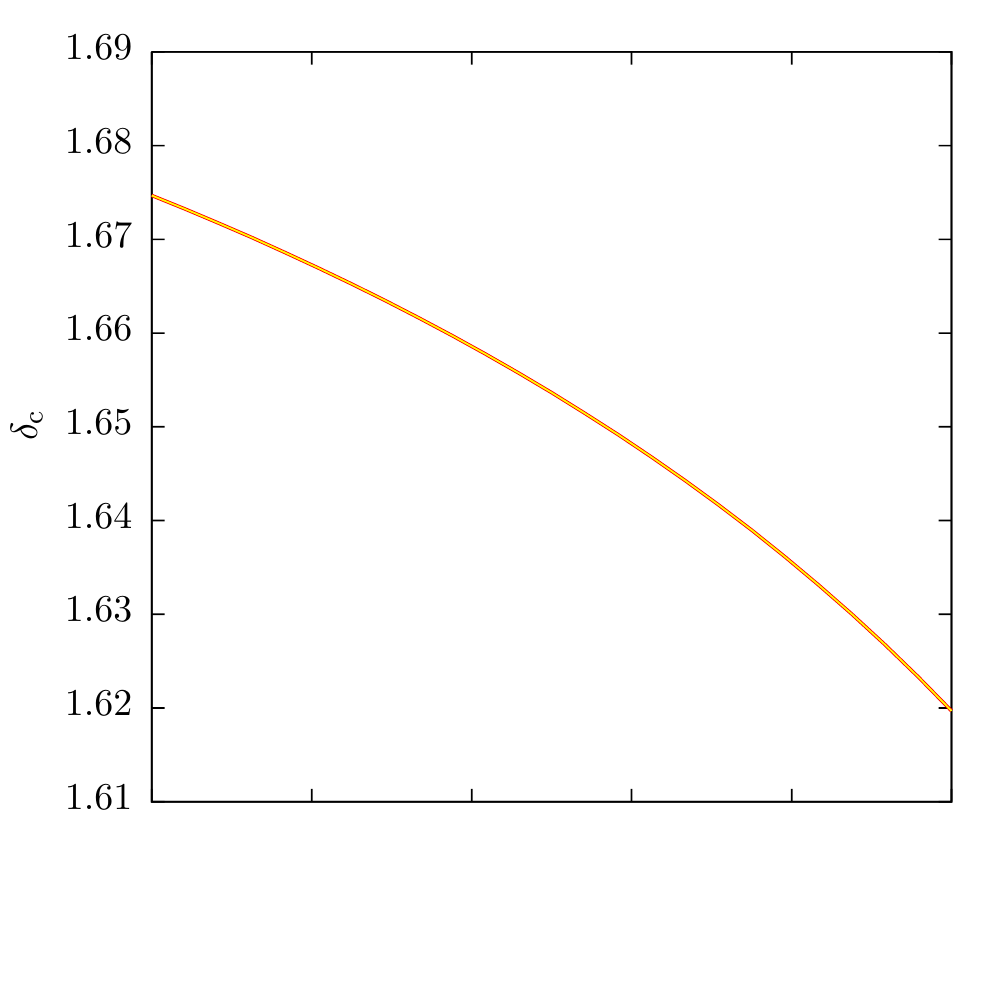}\vspace{-21mm} \\
 \includegraphics[width=80mm]{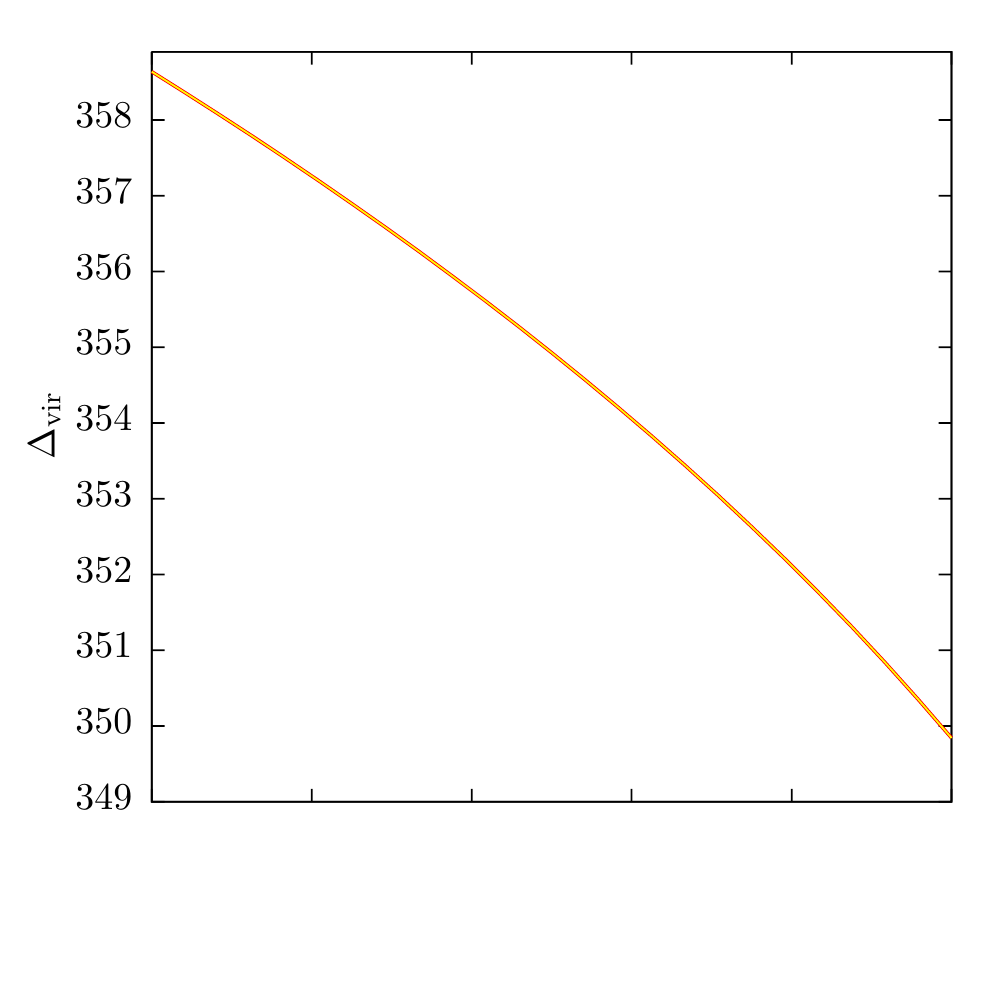}\vspace{-21mm} \\
 \includegraphics[width=80mm]{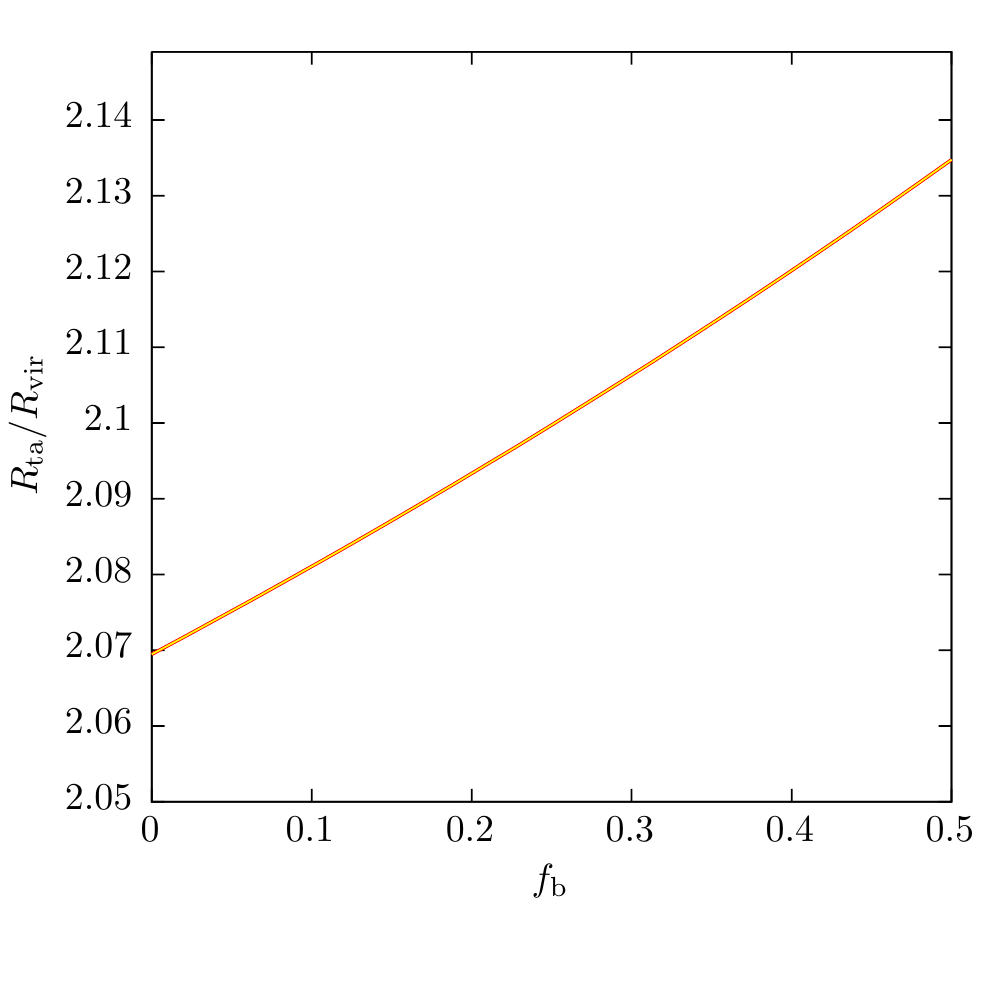}
 \end{tabular}
 \caption{Properties of spherical, nonlinear collapse as a function of the baryon fraction, $f_\mathrm{b}$. The top, middle, and lower panels show the critical overdensity for collapse, the virial density contrast, and the ratio of turnaround to virial radii respectively, all at $z=0$. For the virial density contrast, the line shows the contrast of the collapsed dark matter perturbation relative to the total mean density (i.e. including baryons).}
 \label{fig:nonLinear}
\end{figure}

Figure~\ref{fig:nonLinear} shows the results of the spherical collapse calculations in the limit where baryons do not cluster ($k\gg k_\mathrm{J}$) as a function of baryon fraction. The upper panel shows that the critical overdensity for collapse, $\delta_\mathrm{c}$, decreases slowly as baryon fraction is increased. This does not mean that it is easier for perturbations to collapse when baryons do not participate in clustering---$\delta_\mathrm{c}$ is the amplitude of the corresponding linear perturbation at the time when the non-linear perturbation collapses. Growth of both linear and non-linear perturbations is slowed compared to the dark matter only case when baryons do not cluster, but the linear growth is slowed more, such that the linear perturbation has reached a smaller amplitude by the time of non-linear collapse.

The preceding gives the solution for the limit where baryons do not cluster at all, i.e. $k \rightarrow \infty$. The standard case (treating baryons as a collisionless fluid) gives the opposite limit, $k \rightarrow 0$, where baryon pressure is negligible. To interpolate between the two regimes we make use of the filtering mass \citep{gnedin_probing_1998,naoz_formation_2007}. The fraction of baryons present in a halo of total (i.e. dark matter plus baryonic) mass $M_\mathrm{t}$, is given by
\begin{equation}
 f_\mathrm{b}(M_\mathrm{t},t) = { \Omega_\mathrm{b} / \Omega_0 \over [1+(2^{1/3}-1)8 M_\mathrm{F}(t)/M_\mathrm{t}]^3},
\end{equation}
where $M_\mathrm{F}(t)$ is the filtering mass. Since this is a measure of the extent to which baryons cluster we simply interpolate critical overdensities as
\begin{eqnarray}
 \delta_\mathrm{c}(M_\mathrm{t},t) &=&  \,\,\,\,\, \delta_{\mathrm{c},k \rightarrow 0}(t) f_\mathrm{b}(M_\mathrm{t},t) \nonumber \\
&& + \delta_{\mathrm{c},k\rightarrow\infty}(t) [ 1 - f_\mathrm{b}(M_\mathrm{t},t) ],
\end{eqnarray}
where $\delta_{\mathrm{c},k \rightarrow 0}(t)$, and $\delta_{\mathrm{c},k \rightarrow \infty}(t)$ are the critical overdensities in the $k \rightarrow 0$, and $k \rightarrow \infty$ limits respectively. While this choice of interpolation between the $k\rightarrow 0$ and $k\rightarrow \infty$ regimes is not rigorously justified the difference between $\delta_\mathrm{c}$ in these two regimes is small---for our chosen cosmological model we find that $\delta_\mathrm{c}=1.662\,(1.675)$ for $f_\mathrm{b}=0.167\,(0.000)$---and as such we do not expect the exact choice of how to interpolate to significantly affect our results. We will check this assumption in \S\ref{sec:resultsHaloMassFunction}.

\subsubsection{Virial Density Contrasts}

Although we do not explicitly consider the post-virialization properties of halos in the remainder of this work, for completeness we have calculated the virial density contrast achieved by the collapsing perturbation\footnote{The non-participation of baryons in the collapse of the halo may also affect the internal structure of the halo, such as its concentration. We do not consider this in detail in this work. However, the delayed collapse of the halo would be expected to reduce the concentration. For example, at $z=0$ for a $M=10^7\mathrm{M}_\odot$ halo we will show in \S\protect\ref{sec:results} that $\sigma(M)$ is reduced by a factor of around 8\% relative to a dark matter only model. Using the scaling of concentration with peak height found by \protect\cite{diemer_universal_2014} for low mass halos this would lead to a reduction in concentration of around 8\% for these halos. This effect would be captured by concentration models which explicitly depend on the formation epoch of the halo (e.g. \protect\citealt{ludlow_mass-concentration-redshift_2016}).}. To do this we follow the procedure described by \citeauthor{percival_cosmological_2005}~(\citeyear{percival_cosmological_2005}; \S8). Briefly, the energy of the perturbation is assumed to be conserved between turnaround (when the energy is purely gravitational potential energy, as, by definition, there is no kinetic component at turnaround), and post-virialization (where the energy of the dark matter is shared between gravitational and kinetic in the usual virial ratio). For perturbations below the Jeans scale, the baryonic component does not cluster, and its contribution to the gravitational potential energy of the perturbation therefore differs from the usual case. Following the approach of \cite{percival_cosmological_2005} we find that their eqn.~(38) for the ratio of virial to turnaround radii, $x=R_\mathrm{vir}/R_\mathrm{ta}$, is modified to become
\begin{equation}
 \left( 1 + r - \left[ 1+3w(a_\mathrm{ta}) \right]{q\over 2} \right) x + \left( -{r \over 2 z} \left[ 1+3w(a_\mathrm{vir}) \right]{q\over y}\right) x^3 = {1\over 2},
\end{equation}
where $a_\mathrm{ta}$ and $a_\mathrm{vir}$ are the expansion factors at the epochs of turnaround and virialization respectively,
\begin{eqnarray}
 q &=& {\Omega_\Lambda \over [1+\delta(a_\mathrm{ta})] \Omega_\mathrm{X} (a_\mathrm{ta})}, \nonumber \\
 r &=& {\Omega_\mathrm{b} \over [1+\delta(a_\mathrm{ta})] \Omega_\mathrm{X} (a_\mathrm{ta})}, \nonumber \\
 y &=& {a_\mathrm{ta}^{f(a_\mathrm{ta})} \over a_\mathrm{vir}^{f(a_\mathrm{vir})}}, \nonumber \\
 z &=& {a_\mathrm{ta}^3 \over a_\mathrm{vir}^3},
\end{eqnarray}
measure the contributions of baryons and dark energy to the energy of the perturbation, $f(a)$ depends on the equation of state of dark energy \citep{percival_cosmological_2005}, and $\delta(a_\mathrm{ta})$ is the overdensity of the perturbation at turnaround. We then interpolate between the $k\rightarrow 0$ and $k \rightarrow \infty$ regimes in the same way as for critical overdensities.

The middle panel of figure~\ref{fig:nonLinear} shows the virial density contrast of the collapsed perturbation. Specifically, this is the mean density of the virialized dark matter perturbation relative to the total mean density (i.e. including baryons). The virial density contrast decreases slowly with increasing baryon fraction---the perturbation is more weakly bound as baryons no longer contribute as much to the energy of the collapsing perturbation. For similar reasons, the turnaround radius relative to the virial radius of the perturbation increases slowly with baryon fraction as shown in the lower panel of figure~\ref{fig:nonLinear}. For our chosen cosmological model we find that $\Delta_\mathrm{vir}=356.3\,(358.6)$, and $R_\mathrm{ta}/R_\mathrm{vir}=2.089\,(2.069)$ for $f_\mathrm{b}=0.167\,(0.000)$.

\section{Results}\label{sec:results}

\begin{figure*}
 \begin{tabular}{cc}
  \includegraphics[width=80mm]{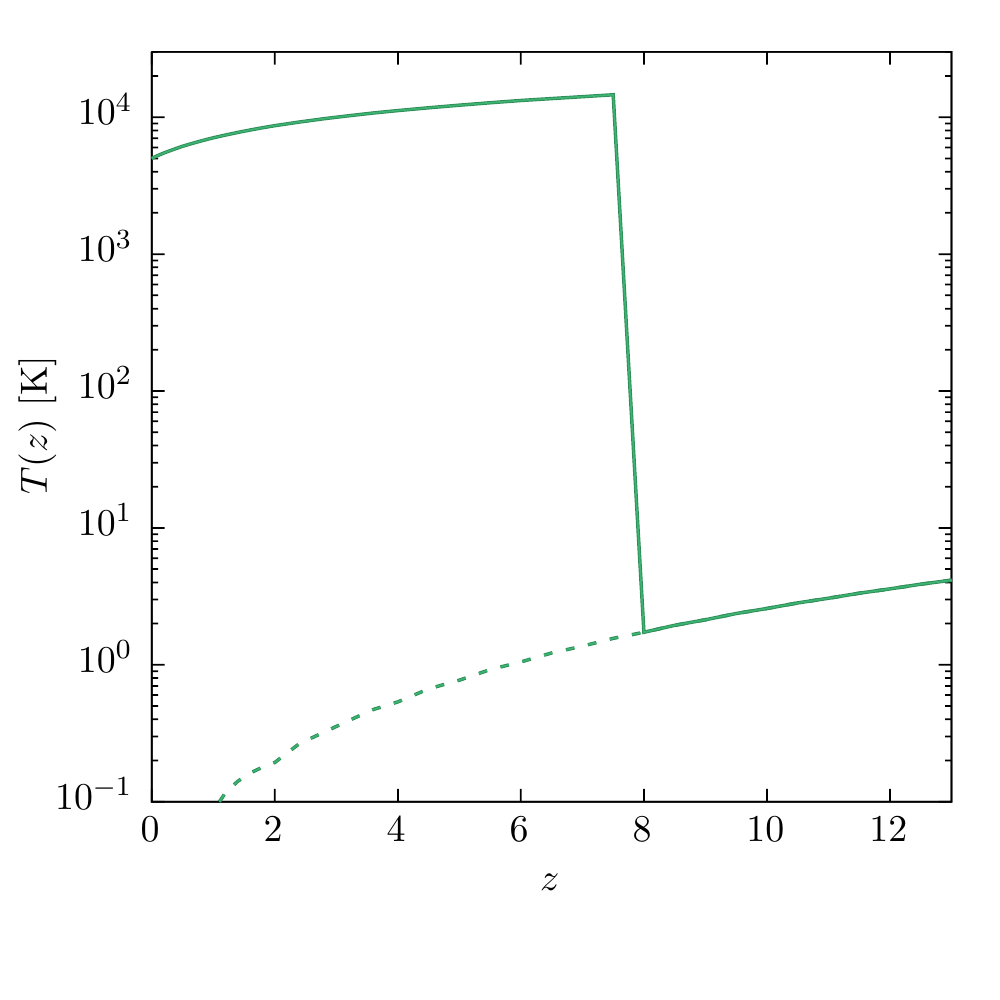} & \includegraphics[width=80mm]{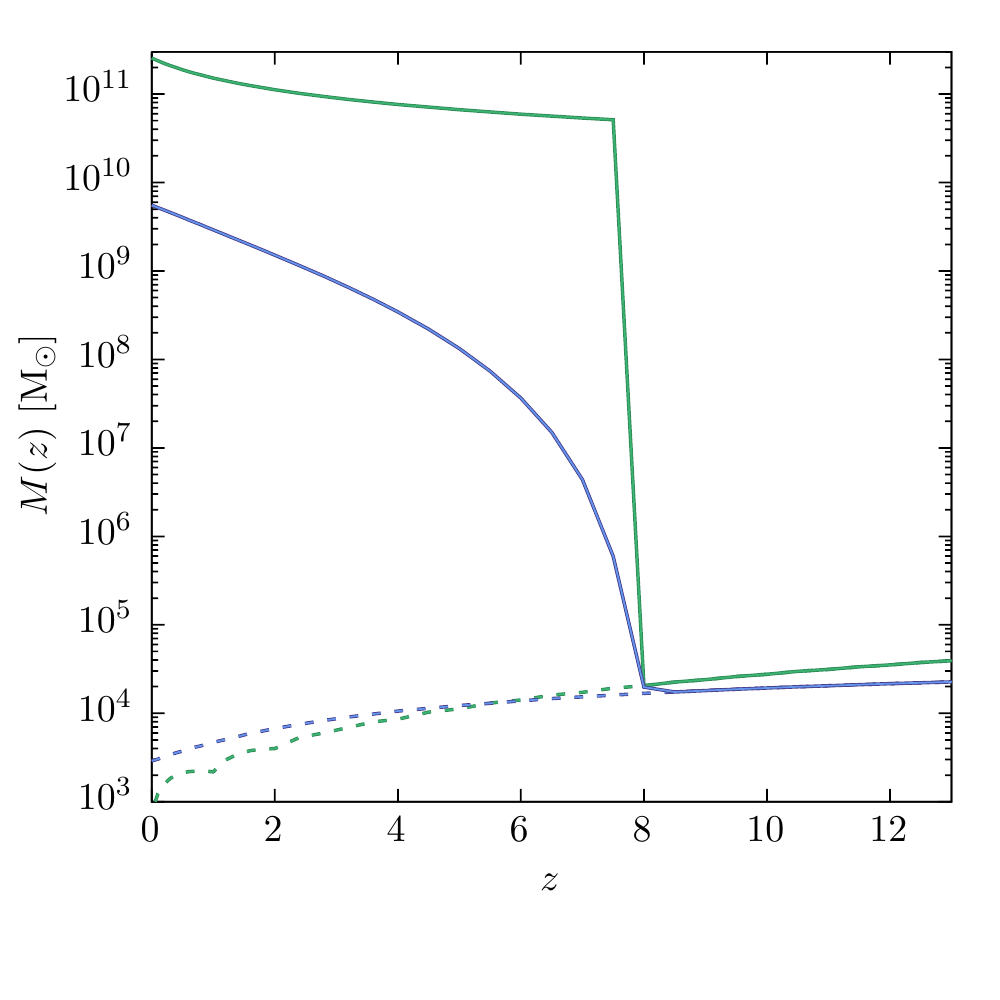} 
 \end{tabular}
 \caption{\emph{Left panel:} The temperature of the IGM as a function of redshift. Dashed lines indicate a model in which the IGM cools adiabatically, while solid lines indicate our simple model of reionization. \emph{Right panel:} The Jeans mass (green lines) and filtering mass (blue lines) as a function of redshift for the same two models.}
 \label{fig:igm}
\end{figure*}

Using the methods developed in the previous section we can now compute halo and subhalo mass functions once we adopt a thermal history for the IGM (which sets the sound speed appearing in equation~\ref{eq:linearGrowth}). We consider three cases: a model in which the IGM cools adiabatically after decoupling from the CMB with no further heating, the thermal history of the ADIAB model of \protect\cite{qin_dark-ages_2017} which includes only shock heating due to structure formation, and a simple reionization model. The thermal evolution of the IGM, along with the Jeans and filtering masses, in the adiabatic and reionization models are shown in Figure~\ref{fig:igm}.

The ADIAB model of \cite{qin_dark-ages_2017} is included so that we may compare our results to those obtained by a full hydrodynamical simulation. We have not shown the results of this model in this section for brevity, but will comment on how well our model agrees with the results of \cite{qin_dark-ages_2017}.

Our simple reionization model assumes that reionization at $z=8$ instantaneously photoheats the IGM to $T=1.5\times 10^4$~K (motivated by the fact that the cooling function of primordial gas increases rapidly at this temperature, making it difficult to heat to much higher temperatures), followed by cooling following a power-law in $(1+z)$ such that the $z=0$ temperature is $T=5 \times 10^3$~K, consistent with the results of \cite{dave_statistical_2001}. This simple model is also in approximate agreement with the measurements of \cite{boera_thermal_2014} at the mean density of the IGM at $z=2.5$ (assuming an exponent of the power-law relation between temperature and density contrast of $\gamma=1.5$ in their analysis). This thermal history is intended to be illustrative of the effects of a likely reionization scenario only---we will comment on the effects of plausible variations to this thermal history in \S\ref{sec:discussion}.

\subsection{Power Spectra}

\begin{figure*}
 \begin{tabular}{cc}
  \includegraphics[width=80mm]{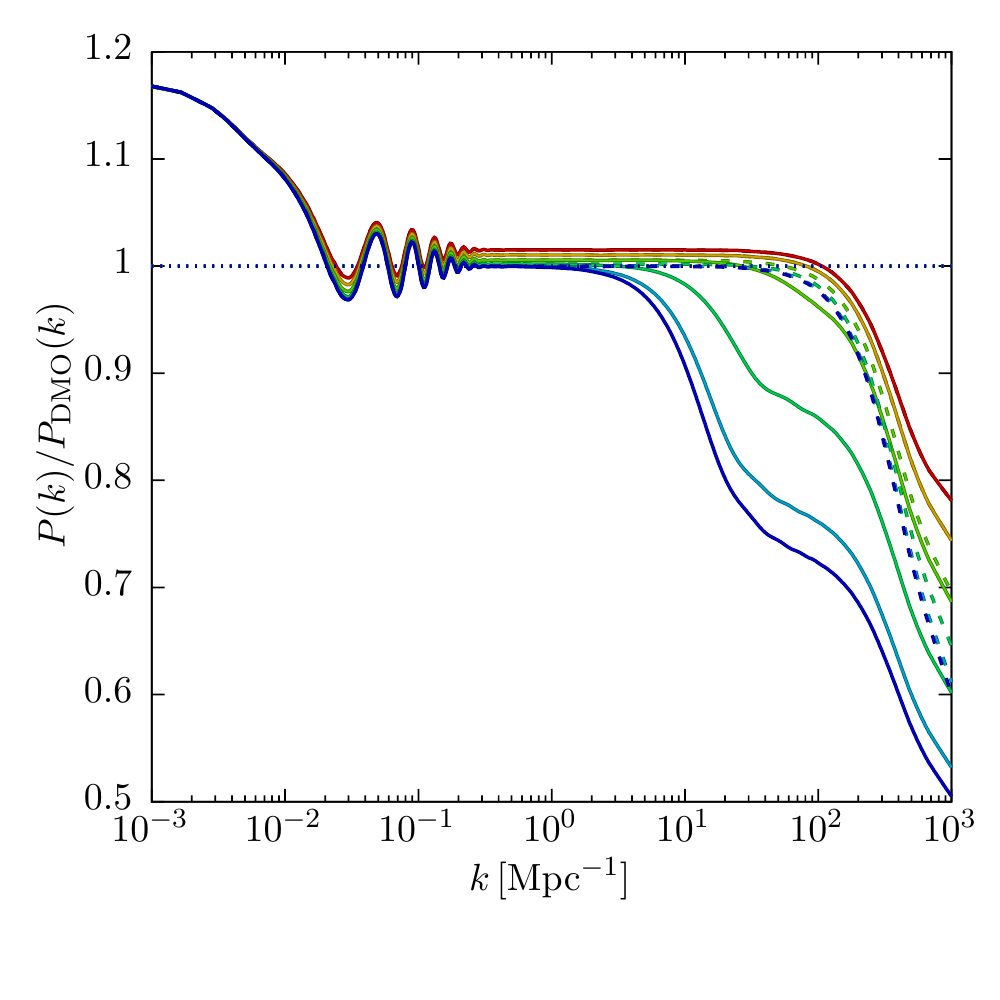} & \includegraphics[width=80mm]{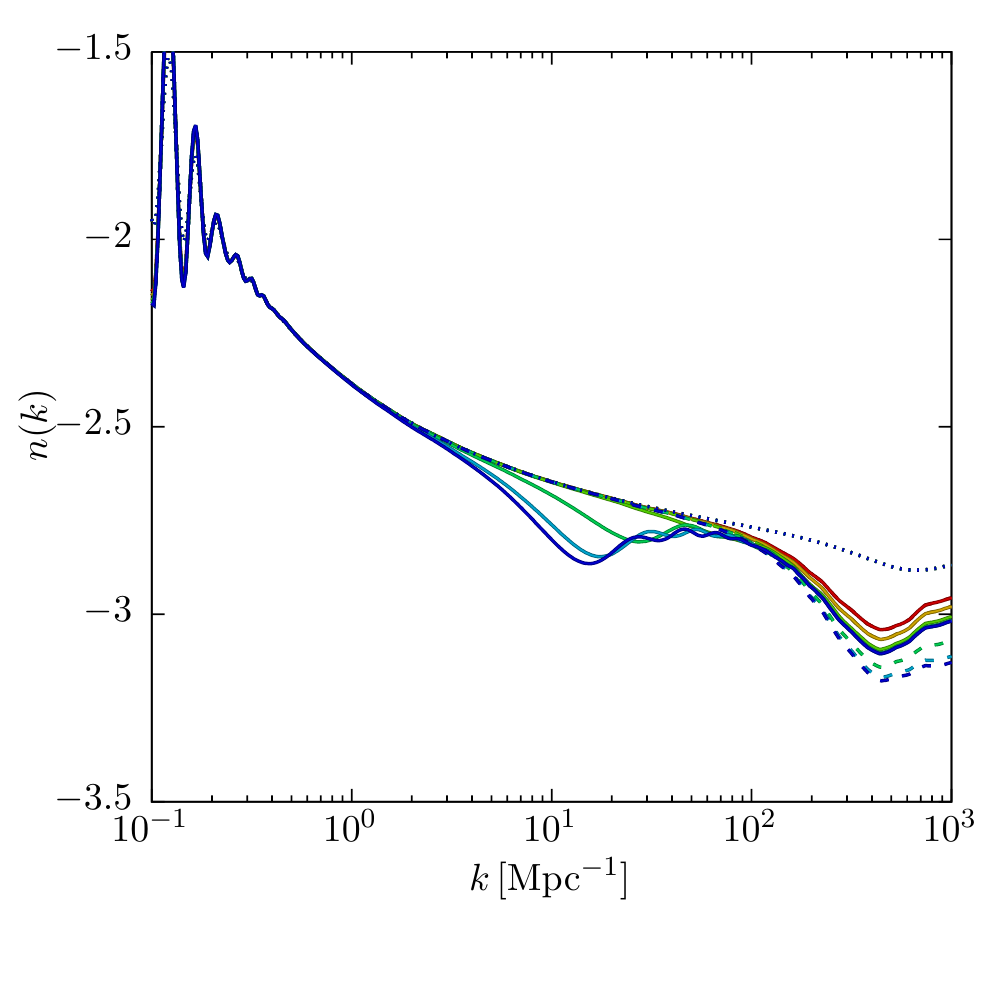} 
 \end{tabular}
 \caption{\emph{Left panel:} The linear theory power spectrum, normalized to the dark matter-only case, at redshifts $z=0.0$, $0.5$, $2.0$, $5.0$, $9.0$, $13.0$ from blue to red. Dotted lines indicate a dark matter only model (i.e. all effects of baryons are ignored), dashed lines indicate a model in which the IGM cools adiabatically, while solid lines indicate our simple model of reionization.\emph{Right panel:} The logarithmic slope of the linear theory power spectrum for the same models and redshifts.}
 \label{fig:powerSpectrum}
\end{figure*}

Figure~\ref{fig:powerSpectrum} shows the linear theory power spectrum normalized to the dark matter-only case (left panel), and the logarithmic slope of the linear theory power spectrum (right panel) at redshifts $z=0.0$, $0.5$, $2.0$, $5.0$, $9.0$, $13.0$ from blue to red. Dotted lines indicate a dark matter only model (i.e. all effects of baryons are ignored), dashed lines indicate a model in which the IGM cools adiabatically, while solid lines indicate our simple model of reionization. Note that dotted and dashed lines coincide almost precisely in this figure. The adiabatic model shows a feature at $k>10^2\,\hbox{Mpc}^{-1}$ which arises from suppression of growth by baryons at very high redshifts. The simple reionization model additionally shows a feature in the power spectrum around $10\,\hbox{Mpc}^{-1}$ which arises from the heating of the IGM in the post-reionization regime\footnote{The wiggles at the high-$k$ regime of this feature are a result of the sharp transition in temperature in the IGM in our simple reionization model. In a more realistic model these features would be smoothed out.}.

\begin{figure}
 \includegraphics[width=80mm]{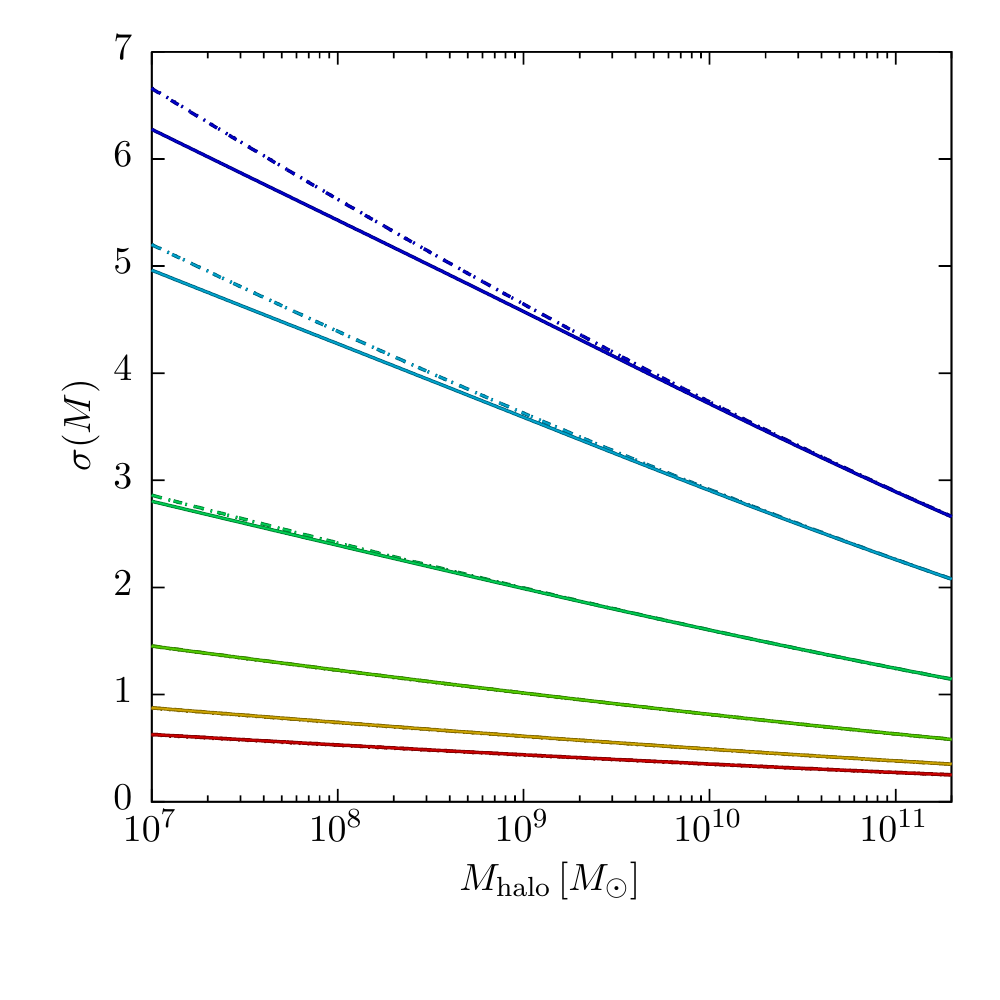}
 \caption{The root-variance as a function of mass, $\sigma(M)$, at redshifts $z=0.0$, $0.5$, $2.0$, $5.0$, $9.0$, $13.0$ from blue to red. Dotted lines indicate a dark matter only model (i.e. all effects of baryons are ignored), dashed lines indicate a model in which the IGM cools adiabatically, while solid lines indicate our simple model of reionization. (Note that dotted and dashed lines coincide in this plot.)}
 \label{fig:massVariance}
\end{figure}

Figure~\ref{fig:massVariance} shows the root-variance of the density field, $\sigma(M)$, in a top-hat filter on mass scale $M$. In the simple reionization model we see that $\sigma(M)$ is reduced at low masses in the post-reionization universe. As such, we may expect the halo mass function to be suppressed on these mass scales also.

\subsection{Halo Mass Functions}\label{sec:resultsHaloMassFunction}

To compute halo mass functions we adopt the approach of \cite{benson_dark_2013}. Briefly, the halo mass function follows the form proposed by \cite{press_formation_1974}, but using a numerical solution to the excursion set barrier crossing problem (which is necessary because the barrier, $\delta_\mathrm{c}$, is no longer independent of mass, and is further modified as proposed by \citealt{sheth_ellipsoidal_2001} to account for ellipsoidal collapse).

We begin by examining our results using the thermal history of the ADIAB model of \cite{qin_dark-ages_2017}---we remind the reader that we do not show results from this model in any figures, but instead simply discuss the results here. Prior to reionization, at $z=9$ and $z=13$, our model predicts no significant change in the halo mass function relative to a pure dark matter model, while \citeauthor{qin_dark-ages_2017}~(\citeyear{qin_dark-ages_2017}; see their Fig.~6) find significantly more suppression in the halo mass function. The lack of suppression in our model at these epochs is not surprising---the temperature of the IGM in the ADIAB model of \cite{qin_dark-ages_2017} remains low ($T<30$~K) until $z\approx 9$. While the Jeans mass is only $M_\mathrm{J}=1.3\times 10^6\mathrm{M}_\odot$ and $M_\mathrm{J}=5.5\times 10^5\mathrm{M}_\odot$ respectively at these epochs, \cite{qin_dark-ages_2017} find suppression of up to 50\% even at masses of $10^9$--$10^{10}\mathrm{M}_\odot$. We hypothesize that the lack of suppression in our model is due to the assumption of a uniform IGM temperature, while in the ADIAB model of \cite{qin_dark-ages_2017} the IGM is heated by virialization shocks which are localized around forming structures, so the effective Jeans mass for collapsing structures may be significantly higher.

By $z=2$, where we may expect the IGM temperature to be much more uniform, the \cite{qin_dark-ages_2017} results show a suppression of the mass function reaching to about 15--20\% at $10^8\mathrm{M}_\odot$. Here we predict a suppression of around 20\% at this same mass and redshift. Our model performs well in matching results from hydrodynamical simulations at late times, where our assumption of a relatively uniform IGM temperature is expected to be most valid.

\begin{figure*}
 \begin{tabular}{cc}
  \includegraphics[width=80mm]{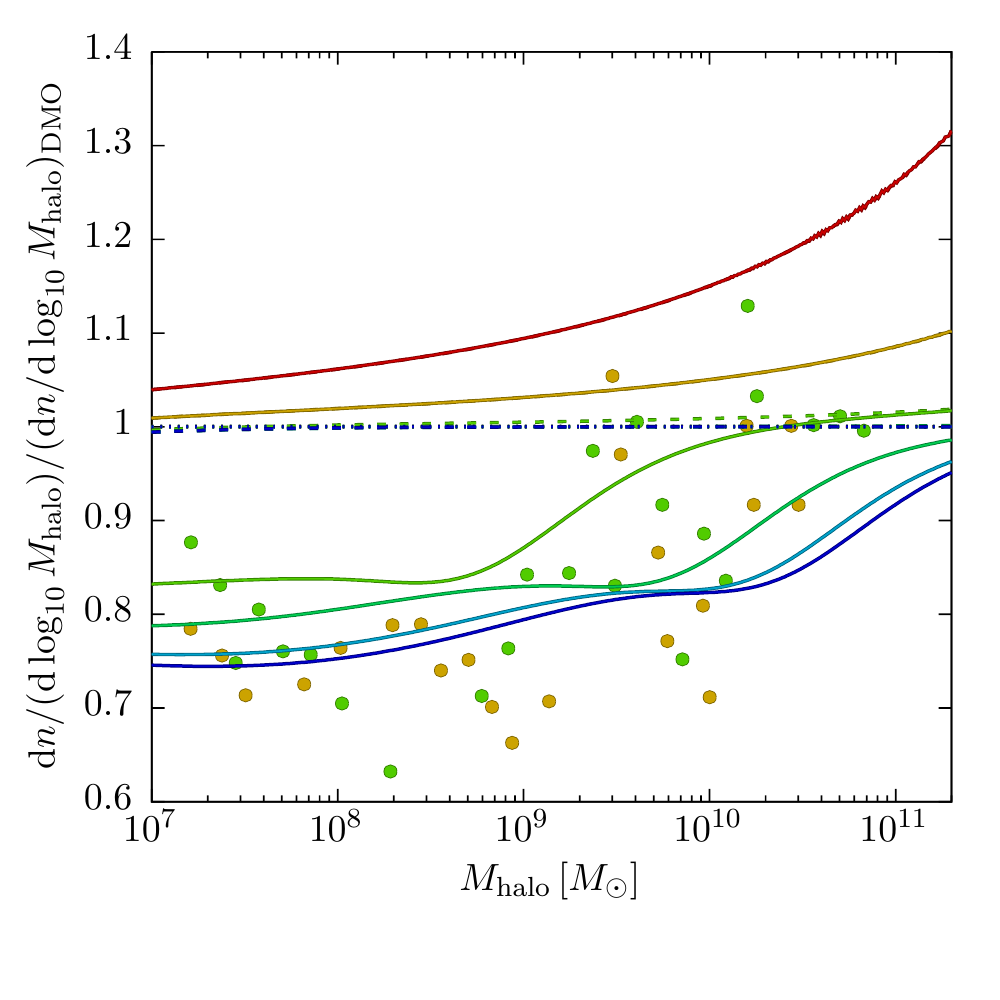} & \includegraphics[width=80mm]{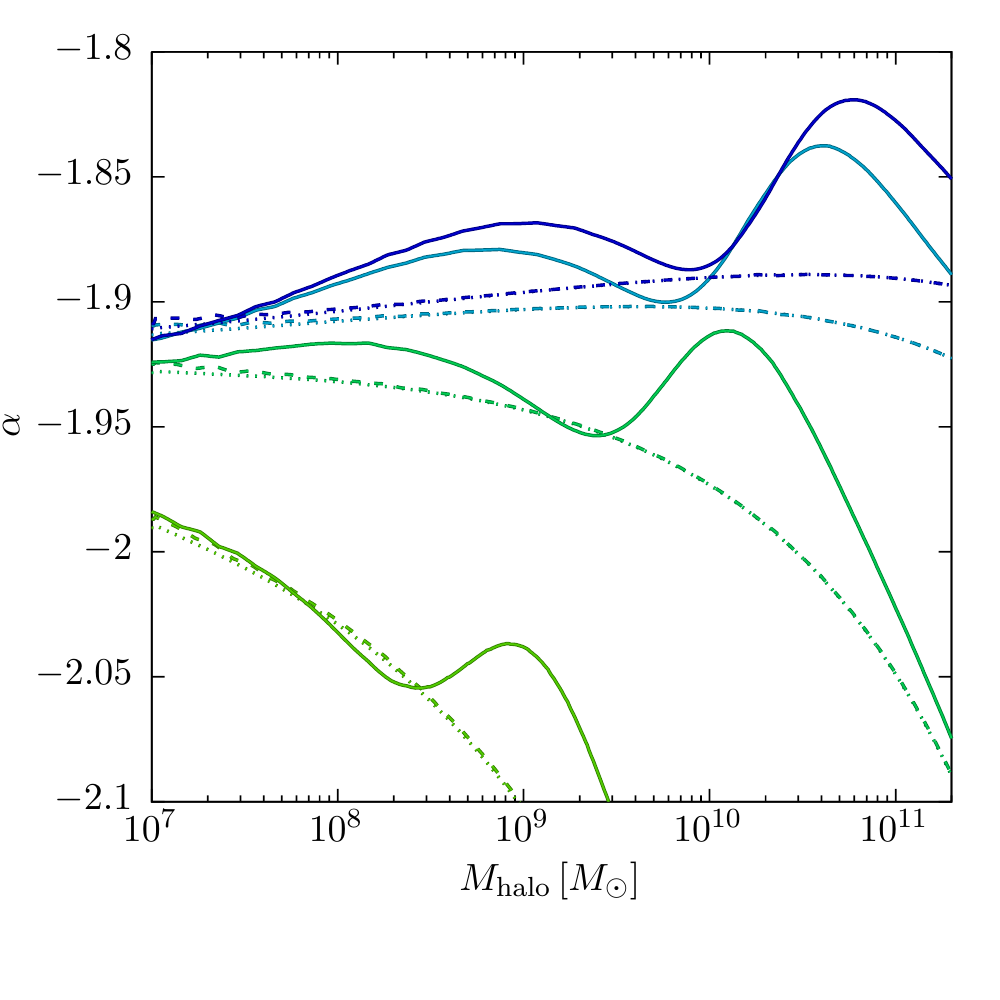}
 \end{tabular}
 \caption{\emph{Left panel:} Halo mass functions at redshifts $z=0.0$, $0.5$, $2.0$, $5.0$, $9.0$, $13.0$ from blue to red, normalized to the dark matter-only case. Dotted lines indicate a dark matter only model (i.e. all effects of baryons are ignored; by construction these lines are horizontal at a value of 1 on the $y$-axis), dashed lines indicate a model in which the IGM cools adiabatically, while solid lines indicate a simple model of reionization. Points show the results of \protect\cite{qin_dark-ages_2017} for their ``{\tt NOSN\_NOZCOOL}'' model at $z=5$ and $z=9$ with colours matched to the corresponding redshift from our model. \emph{Right panel:} Logarithmic slope of the same halo mass functions. Dashed and dotted lines coincide almost precisely in this plot.}
 \label{fig:haloMassFunctionSlopes}
\end{figure*}

We now explore the predictions made by our model for the case where baryons cool adiabatically after they decouple from the CMB, and for simple models of heating of the IGM by reionization. The left panel of figure~\ref{fig:haloMassFunctionSlopes} lines show the resulting halo mass functions at a range of redshifts, normalized to the dark matter only case, while the right panel shows the logarithmic slope of the same mass functions. Prior to the onset of reionization at $z=8$ our adiabatic and reionization models evolve identically, and so the dashed and solid lines coincide precisely at these epochs. In the left panel points show the results of \protect\cite{qin_dark-ages_2017} for their ``{\tt NOSN\_NOZCOOL}'' model (which includes reionization but no other galaxy formation physics and so corresponds most closely to our reioniaztion model) at $z=5$ and $z=9$ with colours matched to the corresponding redshift from our model.

The adiabatically cooling model (dashed lines) actually has a higher halo mass function at $z=9$ and $z=13$ than the dark matter only case. This is because, in the adiabatic case we use initial conditions for perturbations in the dark matter and baryons from CAMB, which includes the full evolution of these perturbations from higher redshifts. As a result, perturbations in baryons are smaller at the initial epoch ($z=150$) than those in the dark matter. In the dark matter only simulations we assume that baryons always behave collisionlessly and so their initial conditions are identical to those of the dark matter. In the adiabatic model therefore, perturbations are suppressed slightly and grow more slowly. Since our power spectra are normalized to a present day $\sigma_8$ this means that at high redshifts the adiabatic model actually had more power than the dark matter only model. In this mass an redshift range the exponential cut-off term in the mass function is important. As such, the mass function is exponentially sensitive to $\sigma(M)$, and so the small increase in power is amplified, leading to a more noticeable increase in the halo mass function.

In our simple reionization model, at epochs post-reionization there is suppression of the mass function, by a level reaching up to 25\% at $M=10^7\mathrm{M}_\odot$ at $z=0$, with suppression beginning to be important at masses of $10^{10}\mathrm{M}_\odot$. We have checked whether our choice of how to interpolate $\delta_\mathrm{c}$ between the $k\rightarrow 0$ and $k\rightarrow \infty$ regimes significantly affects our results by repeating our calculations using a constant $\delta_\mathrm{c}$ fixed at the value expected in each limit. We find that our predicted halo mass function is affected by less than 1\%. The points in the left panel of Fig.~\protect\ref{fig:haloMassFunctionSlopes} show the results from the ``{\tt NOSN\_NOZCOOL}'' model of \protect\cite{qin_dark-ages_2017}. While this model differs from ours in the details of reionization it is nevertheless interesting to compare the qualitative behaviour. At $z=9$ the model of \protect\cite{qin_dark-ages_2017} already shows significant suppression, while ours does not. Similar behaviour was found when comparing to the ADIAB model of \protect\cite{qin_dark-ages_2017}. By $z=5$ however there is qualitative agreement between \protect\cite{qin_dark-ages_2017} and our results, both in the amplitude of the suppression at low masses, and in the mass scale at which the suppression begins to become important.

The right panel of figure~\ref{fig:haloMassFunctionSlopes} shows the logarithmic slope of the same mass functions. Note that dotted and dashed lines coincide almost precisely in this figure. At lower redshifts the dark matter only model has a slope very close to $-1.90$ across the entire range of masses shown. (At higher redshifts the slope is much steeper due to the effects of the exponential cut off in the halo mass function.) In the model including reionization however the slope is seen to be shallower than $-1.90$ across much of this mass range, reaching as high as $\alpha=-1.868$ at $10^9\mathrm{M}_\odot$.

We have examined the effects of changing the assumed temperature evolution of the IGM on these results. We considered a high temperature model in which the IGM is heated to $2.5\times10^4$~K immediately post-reionization, cooling to $10^4$~K at $z=0$ (the upper limit allowed by \citealt{dave_statistical_2001}), and a low temperature model in which the IGM is heated to only $10^4$~K post-reionization, and cools to $2.5\times10^3$~K by $z=0$. The main effect of these changes is to shift the onset of suppression in the mass function (and the corresponding ``bump'' in the slope between $10^{10}$--$10^{11}\hbox{M}_\odot$) to higher or lower mass respectively without substantially changing the magnitude of the suppression or the slope of the mass function at lower masses.

\subsection{Subhalo Mass Function}\label{sec:subhaloMF}

To construct subhalo mass functions we build merger trees. In this case we do not solve the excursion set crossing problem directly ---we find that to compute crossing rates (as needed for merger tree building) for CDM power spectra is prohibitively computationally expensive. Since the barrier, $\delta_\mathrm{c}$ is almost constant (and as shown in \S\ref{sec:resultsHaloMassFunction} treating it as constant makes almost no difference) we simply assume a constant barrier and use the usual solutions for that case---specifically, we use the algorithm of \cite{parkinson_generating_2008} with parameter values from \cite{benson_mass_2017}. We choose host halo masses at $z=0$ in the range $1$--$2\times 10^{13}\mathrm{M}_\odot$ as typical of halo masses of massive elliptical lenses \cite[e.g.][]{gilman_warm_2019}, and a minimum halo mass of $5\times 10^7\mathrm{M}_\odot$ which is below the regime where we expect changes in the slope of the (sub)halo mass function (e.g. see Figure~\ref{fig:haloMassFunctionSlopes}). We allow for subhalos to merge with their host on a dynamical friction timescale as calibrated by \cite{jiang_fitting_2008}, but do not include any other evolution of subhalo masses (such as tidally-induced mass loss). As such, our subhalo mass functions illustrate the effects of the suppression of halo growth by baryons alone, and may not match exactly the results of N-body simulations. Our goal here is to understand the effects of the suppression of growth due to baryons only---effects such as tidal mass loss could be incorporated into our model by following the orbital evolution of each subhalo \citep{taylor_dynamics_2001,pullen_nonlinear_2014}.

\begin{figure*}
 \begin{tabular}{cc}
  \includegraphics[width=80mm]{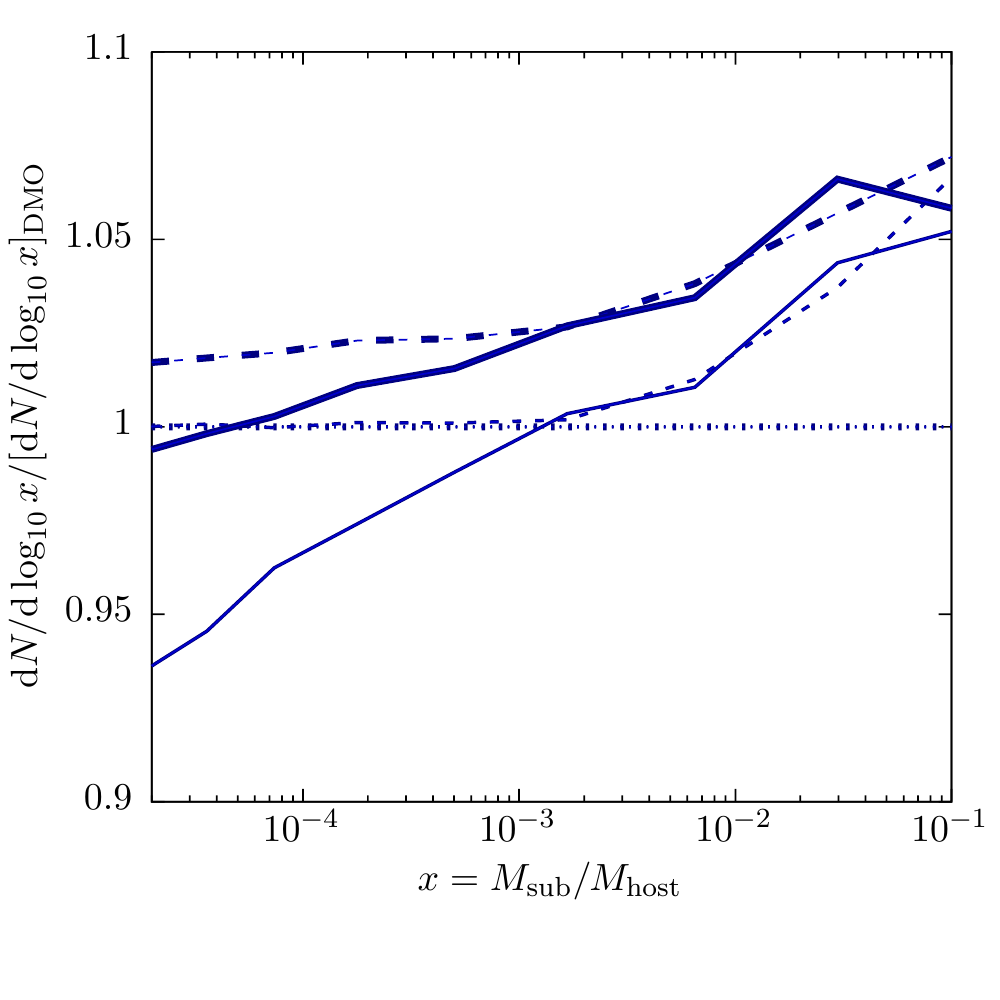} & \includegraphics[width=80mm]{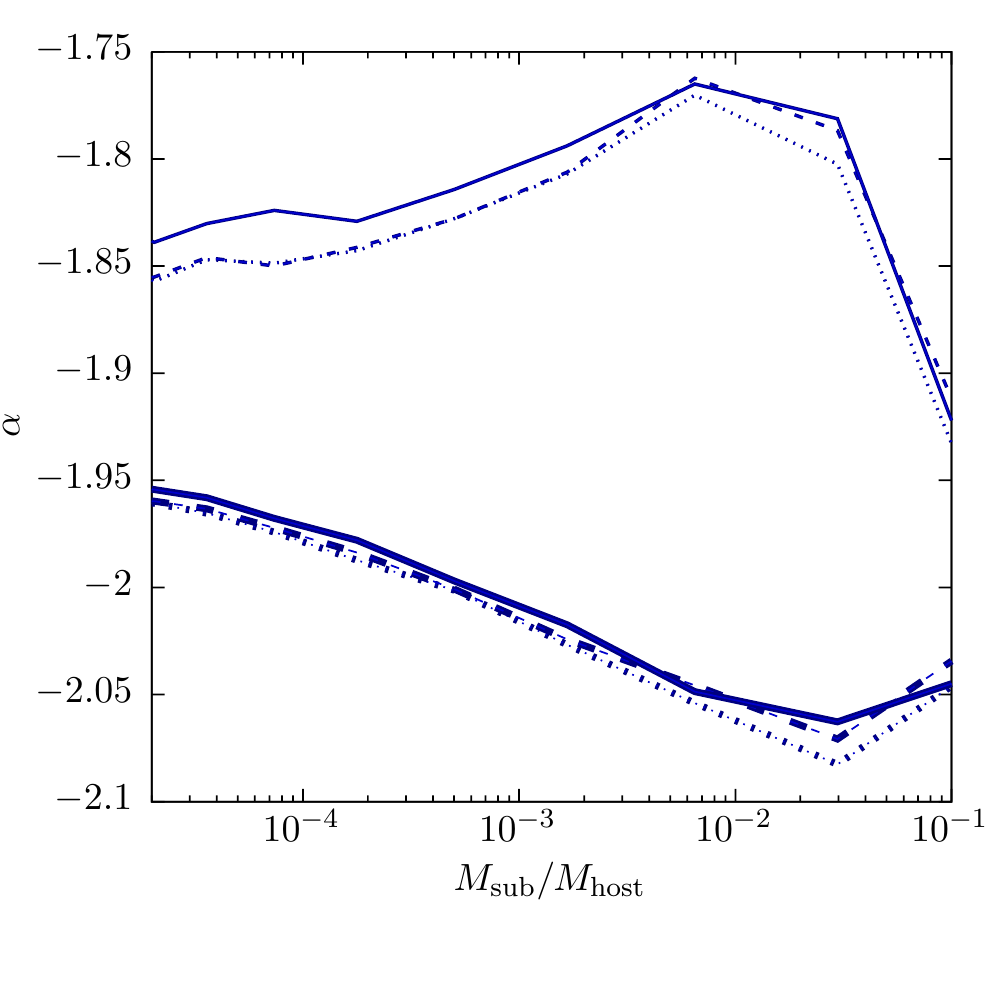}
 \end{tabular}
 \caption{\emph{Left panel:} Subhalo mass functions at redshift $z=0$ for host halo masses in the range $1$--$2\times 10^{13}\mathrm{M}_\odot$ normalized to the dark matter-only case. Dotted lines indicate a dark matter only model (i.e. all effects of baryons are ignored; by construction these lines are horizontal at a value of 1 on the $y$-axis), dashed lines indicate a model in which the IGM cools adiabatically, while solid lines indicate a simple model of reionization. Thick lines indicate the mass function of all subhalos, while thin lines include only those subhalos which accreted into their host halo after $z=1$. \emph{Right panel:} Logarithmic slope of the same subhalo mass functions.}
 \label{fig:subhaloMassFunctionSlopes}
\end{figure*}

Figure~\ref{fig:subhaloMassFunctionSlopes} shows the subhalo mass function for dark matter-only, adiabatic evolution, and our simple reionization model, each normalized to the dark matter only case (left panel), and the slope of the same mass functions (right panel). Thick lines indicate the mass function of all subhalos, while thin lines include only those subhalos which accreted into their host halo after $z=1$. 

Considering first the overall subhalo mass function (thick lines) the adiabatic model shows an increase of a few percent relative to the dark matter-only case. This occurs for the same reason as the increase in the mass function at high redshifts as seen in Figure~\ref{fig:haloMassFunctionSlopes} and discussed in \S\ref{sec:resultsHaloMassFunction}. Since the subhalo mass function is built up by the accretion of subhalos over a wide range of epochs this enhancement is preserved in the $z=0$ subhalo mass function. Relative to the adiabatic model, the subhalo mass function in our simple reionization model is suppressed by around 5\% at the lowest masses---a much weaker suppression than is seen in the mass function (Figure~\ref{fig:haloMassFunctionSlopes}). To understand why the subhalo mass function is less strongly affected by heating of the IGM it is useful to consider the mass function of recently-accreted subhalos---e.g. those accreted after $z=1$ as shown by the thin lines in Figure~\ref{fig:subhaloMassFunctionSlopes}. The subhalo mass function of these recently accreted halos is more strongly affected by baryonic physics at low masses---showing a suppression of up to 10\%. Those subhalos that are accreted early will be less affected by the suppression of structure growth as the Jeans mass is lower at these epochs and small scale perturbations have spent less time below the Jeans scale such that their growth will have been less suppressed. 

To understand the effects of halo mass and redshift on the suppression of the subhalo mass function, we have also computed the subhalo mass function in lower mass ($1$--$2\times 10^{12}\mathrm{M}_\odot$) halos, and at $z=0.5$ (in the original  $1$--$2\times 10^{13}\mathrm{M}_\odot$ halos). We find that the degree of supression changes only weakly as a result of these changes in halo properties---at fixed subhalo mass the suppression in a $\sim10^{12}\mathrm{M}_\odot$ halo at $z=0$ is very similar to that in a $\sim10^{13}\mathrm{M}_\odot$ halo at $z=0$. At $z=0.5$ the suppression is weaker (almost non-existant) for the entire subhalo population of the $\sim10^{13}\mathrm{M}_\odot$ halos, while the suppression of subhalos accreted after $z=1$ is almost the same as for the corresponding $z=0$ halos.

The effects of baryonic physics on the subhalo mass function have been explored by several groups using hydrodynamical simulations. Such simulations should naturally incorporate the same physics that we model in this work, but will also include additional processes which may reduce the number of subhalos at a given mass (e.g. tidal stripping and destruction of subhalos by the galaxy which forms at the center of the host halo---\protect\cite{sawala_shaken_2017} find that baryonic effects suppress subhalo abundances by 20\% at large radii, increasing to 40\% at small radii---indicating that tidal effects are strong at small radii). Therefore, we expect these simulations to predict more suppression than is found in our model. There is remarkable consistency in the results of hydrodynamical simulations for the suppression of low mass subhalos---\cite{brooks_why_2014}, \cite{sawala_bent_2015}, \cite{jahn_dark_2019}, and \cite{samuel_profile_2019} all find that the number of low mass subhalos is suppressed by around 30\% in their hydrodynamical simulations compared to an equivalent dark matter-only simulation. This factor is larger than the 5\% found in this work, indicating (as expected) that additional baryonic effects beyond those considered here are important. \protect\cite{despali_impact_2017} find 20\% and 40\% suppression of the subhalo mass function due to baryonic effects at low masses in the Eagle and Illustris hydrodynamical simulations respectively, indicating that the degree of suppression is dependent on the baryonic physics (which will affect the properties of the central galaxy that contributes to tidal disruption of the subhalos). \protect\cite{chua_subhalo_2017} also find a suppression of around 30\% at low subhalo masses in their analysis of the Illustris simulations, but point out that the abundance of high-mass subahlos is enhanced by the inclusion of baryons (which can act to make the subhalos more strongly bound).

\section{Discussion and Conclusions}\label{sec:discussion}

We have developed a semi-analytic model to describe the effects of baryons on the dark matter halo and subhalo mass functions at scales below the Jeans mass in the IGM. The model agrees well with results from hydrodynamical simulations at late times, but shows less suppression at early times---we hypothesize that this is due to the fact that the IGM is assumed to be uniformly heated in our approach, while in the hydrodynamical simulation to which we compare heating (by virialization shocks) is localized around collapsed halos. The approach developed in this work could plausibly be extended to considering different thermal histories in different environments, potentially allowing a better treatment of the suppression of low mass halo formation at epoch where heating of the IGM is patchy. For the subhalo mass function, the model developed in this work predicts less suppression than is found by hydrodynamical simulations. This is expected as hydrodynamical simulations will naturally incorporate additional sources of suppression (e.g. tidal destruction of subhalos by the massive central galaxy of their host halo) which we do not include here. This indicates that we should combine our model with a detailed treatment of the orbital and tidal evolution of subhalos as was done by \cite{pullen_nonlinear_2014}, but additionally including the effects of baryons. We will leave an examination of these extensions to future papers. 

For simple models of the thermal evolution of the IGM in the post-reionization universe we find that the halo mass function is suppressed by up to 25\% on mass scales below $10^{10}\mathrm{M}_\odot$. The slope of the mass function is also modified on these scales, reaching up to $\alpha=-1.868$ at halo masses of $10^9\mathrm{M}_\odot$. The effects on the subhalo mass function are significantly smaller as a large fraction of the subhalo population formed before the IGM was significantly heated. These mass scales are now beginning to be probed by strong gravitational lensing studies \citep{vegetti_inference_2014,gilman_warm_2019,hsueh_sharp_2019}, which are sensitive to the mass function over a broad range of redshifts from $z=0$ to $z\approx 3$. While \cite{gilman_warm_2019} measured only the subhalo mass function slope (assuming a fixed value for the mass function\footnote{\protect\cite{vegetti_inference_2014} obtained a 95\% lower limit of $\alpha > -2.93$ on the slope of the subhalo mass function, but did not consider any line-of-sight contribution, while \protect\cite{hsueh_sharp_2019} assumed a fix slope of $\alpha = -1.9$.}) given that line of sight halos contribute significantly to the overall lensing cross-section \citep{despali_modelling_2018} we may expect that similarly powerful constraints on the mass function slope could be obtained from this type of observation. Our predictions are consistent with current constraints on the slope and normalization of the mass functions from \protect\cite{vegetti_inference_2014}, and \cite{gilman_warm_2019}, but those constraints can not currently discriminate our prediction from expectations for a pure dark matter model. However, the results of \protect\cite{vegetti_inference_2014} and \cite{gilman_warm_2019} are based on just 11 and 8 strong gravitational lensing systems respectively. As more such systems are obtained and analysed the constraints on the slope and normalization of the (sub)halo mass function are expected to become stronger and may allow the effects of baryons on the dark matter halo mass function to be directly probed.

\section*{Acknowledgements}

We thank Simon Birrer, Xiaolong Du, Daniel Gilman, and Annika Peter for helpful discussions, and the anonymous referee for recommendations which improved this paper. Calculations for this work were carried out on the {\tt mies} compute cluster, made available by a generous grant from the Ahmanson Foundation.

\bibliographystyle{mn2e}
\bibliography{baryonsHaloMassFunction}

\begin{thebibliography}{40}
\expandafter\ifx\csname natexlab\endcsname\relax\def\natexlab#1{#1}\fi

\bibitem[{Benson(2017)}]{benson_mass_2017}
Benson A.~J., 2017, MNRAS, 467, 3454

\bibitem[{Benson {et~al.}(2013)Benson, Farahi, Cole, Moustakas, Jenkins,
  Lovell, Kennedy, Helly, {et~al.}}]{benson_dark_2013}
Benson A.~J., Farahi A., Cole S., Moustakas L.~A., Jenkins A., Lovell M.,
  Kennedy R., Helly J., {et~al.}, 2013, MNRAS, 428, 1774

\bibitem[{Boera {et~al.}(2014)Boera, Murphy, Becker, \&
  Bolton}]{boera_thermal_2014}
Boera E., Murphy M.~T., Becker G.~D., Bolton J.~S., 2014, MNRAS, 441, 1916

\bibitem[{Bond {et~al.}(1991)Bond, Cole, Efstathiou, \&
  Kaiser}]{bond_excursion_1991}
Bond J.~R., Cole S., Efstathiou G., Kaiser N., 1991, ApJ, 379, 440

\bibitem[{Bower(1991)}]{bower_evolution_1991}
Bower R.~G., 1991, MNRAS, 248, 332

\bibitem[{Brooks \& Zolotov(2014)}]{brooks_why_2014}
Brooks A.~M., Zolotov A., 2014, ApJ, 786, 87

\bibitem[{Challinor \& Lewis(2011)}]{challinor_linear_2011}
Challinor A., Lewis A., 2011, Physical Review D, 84, 043516

\bibitem[{Chua {et~al.}(2017)Chua, Pillepich, Rodriguez-Gomez, Vogelsberger,
  Bird, \& Hernquist}]{chua_subhalo_2017}
Chua K. T.~E., Pillepich A., Rodriguez-Gomez V., Vogelsberger M., Bird S.,
  Hernquist L., 2017, MNRAS, 472, 4343

\bibitem[{Dav\'e \& Tripp(2001)}]{dave_statistical_2001}
Dav\'e R., Tripp T.~M., 2001, ApJ, 553, 528

\bibitem[{Despali \& Vegetti(2017)}]{despali_impact_2017}
Despali G., Vegetti S., 2017, MNRAS, 469, 1997

\bibitem[{Despali {et~al.}(2018)Despali, Vegetti, White, Giocoli, \& van~den
  Bosch}]{despali_modelling_2018}
Despali G., Vegetti S., White S. D.~M., Giocoli C., van~den Bosch F.~C., 2018,
  MNRAS, 475, 5424

\bibitem[{Diemer \& Kravtsov(2014)}]{diemer_universal_2014}
Diemer B., Kravtsov A.~V., 2014, arXiv:1407.4730 [astro-ph], arXiv: 1407.4730

\bibitem[{Fiacconi {et~al.}(2016)Fiacconi, Madau, Potter, \&
  Stadel}]{fiacconi_cold_2016}
Fiacconi D., Madau P., Potter D., Stadel J., 2016, ApJ, 824, 144

\bibitem[{Gilman {et~al.}(2019)Gilman, Birrer, Nierenberg, Treu, Du, \&
  Benson}]{gilman_warm_2019}
Gilman D., Birrer S., Nierenberg A., Treu T., Du X., Benson A., 2019, arXiv
  e-prints, arXiv:1908.06983

\bibitem[{Gnedin(2000)}]{gnedin_effect_2000}
Gnedin N.~Y., 2000, ApJ, 542, 535

\bibitem[{Gnedin \& Hui(1998)}]{gnedin_probing_1998}
Gnedin N.~Y., Hui L., 1998, MNRAS, 296, 44

\bibitem[{Gunn \& Gott(1972)}]{gunn_infall_1972}
Gunn J.~E., Gott J.~R., 1972, ApJ, 176, 1

\bibitem[{Hsueh {et~al.}(2019)Hsueh, Enzi, Vegetti, Auger, Fassnacht, Despali,
  Koopmans, \& McKean}]{hsueh_sharp_2019}
Hsueh J.-W., Enzi W., Vegetti S., Auger M.~W., Fassnacht C.~D., Despali G.,
  Koopmans L. V.~E., McKean J.~P., 2019, MNRAS

\bibitem[{Hu \& Eisenstein(1998)}]{hu_small-scale_1998}
Hu W., Eisenstein D.~J., 1998, ApJ, 498, 497

\bibitem[{Jahn {et~al.}(2019)Jahn, Sales, Wetzel, Boylan-Kolchin, Chan,
  El-Badry, Lazar, \& Bullock}]{jahn_dark_2019}
Jahn E.~D., Sales L.~V., Wetzel A., Boylan-Kolchin M., Chan T.~K., El-Badry K.,
  Lazar A., Bullock J.~S., 2019, MNRAS, 2117

\bibitem[{Jiang {et~al.}(2008)Jiang, Jing, Faltenbacher, Lin, \&
  Li}]{jiang_fitting_2008}
Jiang C.~Y., Jing Y.~P., Faltenbacher A., Lin W.~P., Li C., 2008, ApJ, 675,
  1095

\bibitem[{Komatsu {et~al.}(2011)Komatsu, Smith, Dunkley, Bennett, Gold,
  Hinshaw, Jarosik, Larson, {et~al.}}]{komatsu_seven-year_2011}
Komatsu E., Smith K.~M., Dunkley J., Bennett C.~L., Gold B., Hinshaw G.,
  Jarosik N., Larson D., {et~al.}, 2011, ApJS, 192, 18

\bibitem[{Ludlow {et~al.}(2016)Ludlow, Bose, Angulo, Wang, Hellwing, Navarro,
  Cole, \& Frenk}]{ludlow_mass-concentration-redshift_2016}
Ludlow A.~D., Bose S., Angulo R.~E., Wang L., Hellwing W.~A., Navarro J.~F.,
  Cole S., Frenk C.~S., 2016, MNRAS, 460, 1214

\bibitem[{Naoz \& Barkana(2007)}]{naoz_formation_2007}
Naoz S., Barkana R., 2007, MNRAS, 377, 667

\bibitem[{Parkinson {et~al.}(2008)Parkinson, Cole, \&
  Helly}]{parkinson_generating_2008}
Parkinson H., Cole S., Helly J., 2008, MNRAS, 383, 557

\bibitem[{Peebles(1980)}]{peebles_large-scale_1980}
Peebles P. J.~E., 1980, Large-Scale Structure of the Universe by Phillip James
  Edwin Peebles. Princeton University Press

\bibitem[{Percival(2005)}]{percival_cosmological_2005}
Percival W.~J., 2005, Astronomy and Astrophysics, 443, 819

\bibitem[{Percival {et~al.}(2000)Percival, Miller, \&
  Peacock}]{percival_analytic_2000}
Percival W.~J., Miller L., Peacock J.~A., 2000, MNRAS, 318, 273

\bibitem[{Press \& Schechter(1974)}]{press_formation_1974}
Press W.~H., Schechter P., 1974, ApJ, 187, 425

\bibitem[{Pullen {et~al.}(2014)Pullen, Benson, \&
  Moustakas}]{pullen_nonlinear_2014}
Pullen A.~R., Benson A.~J., Moustakas L.~A., 2014, ApJ, 792, 24

\bibitem[{Qin {et~al.}(2017)Qin, Duffy, Mutch, Poole, Geil, Angel, Mesinger, \&
  Wyithe}]{qin_dark-ages_2017}
Qin Y., Duffy A.~R., Mutch S.~J., Poole G.~B., Geil P.~M., Angel P.~W.,
  Mesinger A., Wyithe J. S.~B., 2017, MNRAS, 467, 1678

\bibitem[{Samuel {et~al.}(2019)Samuel, Wetzel, Tollerud, Garrison-Kimmel,
  Loebman, El-Badry, Hopkins, Boylan-Kolchin, {et~al.}}]{samuel_profile_2019}
Samuel J., Wetzel A., Tollerud E., Garrison-Kimmel S., Loebman S., El-Badry K.,
  Hopkins P.~F., Boylan-Kolchin M., {et~al.}, 2019, arXiv e-prints,
  arXiv:1904.11508

\bibitem[{Sawala {et~al.}(2015)Sawala, Frenk, Fattahi, Navarro, Bower, Crain,
  Dalla~Vecchia, Furlong, {et~al.}}]{sawala_bent_2015}
Sawala T., Frenk C.~S., Fattahi A., Navarro J.~F., Bower R.~G., Crain R.~A.,
  Dalla~Vecchia C., Furlong M., {et~al.}, 2015, MNRAS, 448, 2941

\bibitem[{Sawala {et~al.}(2017)Sawala, Pihajoki, Johansson, Frenk, Navarro,
  Oman, \& White}]{sawala_shaken_2017}
Sawala T., Pihajoki P., Johansson P.~H., Frenk C.~S., Navarro J.~F., Oman
  K.~A., White S. D.~M., 2017, MNRAS, 467, 4383

\bibitem[{Schaller(2015)}]{schaller_effects_2015}
Schaller M., 2015

\bibitem[{Sheth {et~al.}(2001)Sheth, Mo, \& Tormen}]{sheth_ellipsoidal_2001}
Sheth R.~K., Mo H.~J., Tormen G., 2001, MNRAS, 323, 1

\bibitem[{Springel {et~al.}(2008)Springel, Wang, Vogelsberger, Ludlow, Jenkins,
  Helmi, Navarro, Frenk, {et~al.}}]{springel_aquarius_2008}
Springel V., Wang J., Vogelsberger M., Ludlow A., Jenkins A., Helmi A., Navarro
  J.~F., Frenk C.~S., {et~al.}, 2008, MNRAS, 391, 1685

\bibitem[{Taylor \& Babul(2001)}]{taylor_dynamics_2001}
Taylor J.~E., Babul A., 2001, ApJ, 559, 716

\bibitem[{Vegetti {et~al.}(2014)Vegetti, Koopmans, Auger, Treu, \&
  Bolton}]{vegetti_inference_2014}
Vegetti S., Koopmans L. V.~E., Auger M.~W., Treu T., Bolton A.~S., 2014, MNRAS,
  442, 2017

\bibitem[{Weinberg \& Kamionkowski(2003)}]{weinberg_constraining_2003}
Weinberg N.~N., Kamionkowski M., 2003, MNRAS, 341, 251

\end{thebibliography}

\end{document}